\documentclass[twocolumn,showpacs,preprintnumbers,amsmath,amssymb]{revtex4}
\usepackage{graphicx}
\usepackage{dcolumn}
\usepackage{bm}
\usepackage{multirow}
\usepackage{tabularx}
\begin{document}

\title{Re-evaluation of the $^{22}$Ne($p$,$\gamma$)$^{23}$Na reaction rate: $R-$matrix analysis of the non-resonant capture and effect of the 8945 keV (${7/2}^{-}$) resonance strength }




\author{Sk Mustak Ali$^1$}
\email{mustak21.mail@gmail.com}
\altaffiliation{Present Address: Facility for Rare Isotope Beam, Michigan State University, East Lansing, Michigan 48824, USA}
\author{Rajkumar Santra$^2$}
\altaffiliation{Corresponding author}
\email{rajkumarsantra2013@gmail.com}
\author{Sathi Sharma$^3$}
\author{Ashok kumar Mondal$^4$}

\affiliation{$^1$Department of Physics, Bose Institute, 93/1 APC Road, Kolkata 700009, India}
\affiliation{$^2$Variable Energy Cyclotron Centre, 1/AF, Bidhan Nagar, Kolkata 700064, India}
\affiliation{$^3$Department of Physics, Manipal Institute of Technology, India}
\affiliation{$^4$Department of Physics, Manipal University Jaipur}

\date{\today}
\begin{abstract}
The $^{22}$Ne($p,\gamma$)$^{23}$Na capture reaction is a key member of the Ne-Na cycle of hydrogen burning. The rate of this reaction is critical in classical novae nucleosynthesis and hot bottom burning processes (HBB) in asymptotic giant branch (AGB) stars.  Despite its astrophysical importance, significant uncertainty remains in the reaction rate due to several narrow low energy resonances lying near the Gamow window. The present work revisits this reaction by examining the contribution of the 8664 keV subthreshold state and the 151 keV doublet resonance state of 7/2$^-$ configuration in $^{23}$Na. Finite range distorted-wave Born approximation (FRDWBA) analyses of existing $^{22}$Ne($^3$He,$d$)$^{23}$Na transfer reaction data were carried out to extract the peripheral asymptotic normalization coefficients (ANC) of the 8664 keV state. The ANC value obtained in the present work is $\sim 25\%$ higher compared to the previous work by Santra et al.~\cite{SA20}.  Systematic $R$-matrix calculations were performed to obtain the non-resonant astrophysical $S$-factor utilizing the enhanced ANC value. The resonance strengths of the 8945 keV doublets were deduced from shell model calculations. The total reaction rate is found to be $\sim 15\%$ higher at temperatures relevant for the HBB processes, compared to the recent rate measured by Williams et al.~\cite{WI20}, and matches the rate by Williams et al.~\cite{WI20} at temperatures of interest for classical novae nucleosynthesis.

\end{abstract}
 
\maketitle

\section{Introduction}
The neon-sodium (Ne-Na) cycle is of enormous importance in stellar nucleosynthesis as it is responsible for the hydrogen burning in massive stars, and is involved in the synthesis of elements between Ne and Mg~\cite{MA57,RO75}. Within the Ne-Na cycle, the proton capture reaction $^{22}$Ne($p,\gamma$)$^{23}$Na ($Q=8794.11 \pm 0.02$ keV) is of significant interest. It not only consumes $^{22}$Ne, which is the third most abundant nuclide produced in stellar helium burning, but also produces $^{23}$Na, the only stable isotope of sodium~\cite{BU06,KA11}. 
This reaction influences the weak $s-$process nucleosynthesis by competing with the $^{22}$Ne($\alpha$,$n$)$^{25}$Mg reaction, which is a major neutron source in asymptotic giant branch (AGB) stars. 
The rate of this reaction impacts the stellar models that seek to explain the puzzling anticorrelation in oxygen and sodium abundances observed in
globular clusters~\cite{GR12,VE18}. It affects the abundance ratios of Ne isotopes in presolar grains extracted from meteorites~\cite{LE20}. Further, sensitivity studies have shown that the nuclear uncertainties of the $^{22}$Ne($p,\gamma$)$^{23}$Na reaction can have drastic impact on the $^{22}$Ne and $^{23}$Na abundances in classical novae nucleosynthesis~\cite{IL02}.    

The reaction rate of $^{22}$Ne($p,\gamma$)$^{23}$Na at the
astrophysical energies depends on the contribution of several low energy resonances in $^{23}$Na and a slowly varying non-resonant capture component.
The uncertainty in the rate spanned a factor of 1000 between the rates from NACRE~\cite{AN99} and others~\cite{HA01,IL10,SA13}. The dominant source of this uncertainty is due to various unobserved or poorly constrained narrow resonances at the relevant astrophysical energies.
In recent years, several measurements were carried out to address the large discrepancy in the $^{22}$Ne($p,\gamma$)$^{23}$Na rate by precisely measuring the pertinent resonance strengths at proton energies $E_{p} \sim 70 - 500$ keV~\cite{CA15,DE16,KE17,CA18,FE18,WI20}. As a result, the uncertainty in the $^{22}$Ne($p,\gamma$)$^{23}$Na rate was reduced by 3 orders of magnitude at $T=0.1$ GK~\cite{FE18}. 

Despite this major improvement, contentions remain on the resonance strength measurements and existence of some of the low energy resonances lying inside and near the Gamow window. The 8945 keV state in $^{23}$Na affects the $^{22}$Ne($p,\gamma$)$^{23}$Na reaction rate as it lies within the Gamow window at $T=0.1$ GK. Previous measurements of the 8945 keV resonance considered it as a single state with $J^{\pi}={7/2}^{-}$~\cite{GO82,HA01}.
However, the measurement by Jenkins~\textit{et~al.}~\cite{JE13} with Gammasphere reported that this resonance actually comprises a doublet, one with $J^\pi = {7/2}^{-}$ and the other with a tentative $J^{\pi}={3/2}^{+}$. Several direct measurements have obtained the resonance strength for the ${3/2}^{+}$ state, and the value ranges from $1.48 \times 10^{-7}$ eV to $2.2 \times 10^{-7}$ eV~\cite{DE16,KE17,CA18,FE18,WI20}. But no such direct measurements exist for the ${7/2}^{-}$ state and only an upper limit of $9.7 \times 10^{-8}$ eV has been recommended for its strength from ($^3$He,$d$) transfer reaction~\cite{HA01}. This strength was calculated from the spectroscopic factor for the 8945 keV state, assuming $l = 3$ transfer, consistent with
$J^{\pi}={7/2}^{-}$. However, this assumption is questionable due to the limited number of data points in the angular distribution of the 8945 keV state.

Recently, Santra et al.~\cite{SA20} reanalyzed the data of Hale et al.~\cite{HA01} considering the contribution of both the $3/2^{+}$ ($l=2$ transfer) and $7/2^{-}$ ($l=3$ transfer) states, extracting the spectroscopic factors for both the states. 
They observed a better reproduction of the limited angular distribution data. In their indirect study of the $^{22}$Ne($p,\gamma$)$^{23}$Na reaction, they carried out a systematic $R-$matrix analysis of the direct capture (DC) component and including the contribution of the 8664 keV subthreshold state in $^{23}$Na.
The low energy behaviour of the $S-$factor of $^{22}$Ne($p,\gamma$)$^{23}$Na reaction is controlled by this subthreshold resonance~\cite{FE18}.  In Ref.~\cite{SA20}, the $R-$matrix calculations were constrained by the asymptotic normalization coefficients (ANC) extracted from the $^{20}$Ne($^{3}$He,$d$)$^{23}$Na transfer data at 15~MeV~\cite{PO71} for the first six bound states and 20~MeV~\cite{HA01} for the 8664 keV state. However, the ANC value of $144$ fm$^{-1/2}$ obtained for the 8664 keV state in Ref.~\cite{SA20} did not satisfy the necessary peripherality conditions. The resulting $S-$factor using this ANC value could not reproduce the DC $\rightarrow$ 8664 keV capture data, particularly for $E_{p}<500$ keV. A better fit to the data was obtained by simultaneous R-matrix fit to the direct capture data of Rolfs~\textit{et~al.}~\cite{RO75}, Gorres~\textit{et~al.}~\cite{GO82} and Ferraro~\textit{et~al.}~\cite{FE18} keeping the ANC and the $\Gamma_\gamma$ values of the background poles as free parameters. 
As a result, they could reproduce the rising effect in the low energy astrophysical $S-$factor of the ground state capture data as observed by Ferraro~\textit{et~al.}~\cite{FE18}. They reported a value of $48.8\pm 9.5$~keV~b for the total direct capture $S-$factor at zero relative energy, and the resultant reaction rate was distinctly higher compared to the previously obtained rates~\cite{HA01,KE17,FE18} for $T \leq 0.1$~GK.  

In the present work, we attempt to re-examine the $^{22}$Ne($p,\gamma$)$^{23}$Na reaction by  
focusing on the extraction of peripheral ANC of the 8664 keV state. The angular distribution data of the $^{20}$Ne($^{3}$He,$d$)$^{23}$Na one-proton stripping reaction at energies of 12 and 15 MeV has been used to obtain the corresponding ANC and its peripheral nature is checked. Also, the contributions of the excited states 7080, 7449 and 7890 keV which were not considered in the previous analysis~\cite{SA20} have been included in the present work. 
As discussed earlier, the spectroscopic factor for the ${7/2}^{-}$ configuration of the 8945 keV state still remain elusive of direct measurements. Hence, detailed microscopic shell model calculations have been performed to yield the required proton width ($\Gamma_p$) and the corresponding resonance strength for this state. The resultant
reaction rate  as a function of temperature is compared with the recent
measurement by Williams et al.~\cite{WI20}.

\section{Analysis}

\subsection{Finite-range DWBA analysis and ANC extraction}
The finite-range distorted wave Born approximation (FRDWBA) calculations were performed for the 8664~keV (${1/2}^+$) subthreshold state in $^{23}$Na using the existing angular distribution data of  $^{22}$Ne($^3$He,$d$)$^{23}$Na reaction at bombarding energy of 15 MeV~\cite{PO71}.
The FRDWBA calculations required the optical model potential (OMP) parameters for the entrance channel $^{22}$Ne+$^3$He, 
 exit channel $d$+$^{23}$Na, and the core–core $^{22}$Ne+$d$ interactions.
\begin{figure}[h!]
    \centering 
    \includegraphics[width =0.5\textwidth]{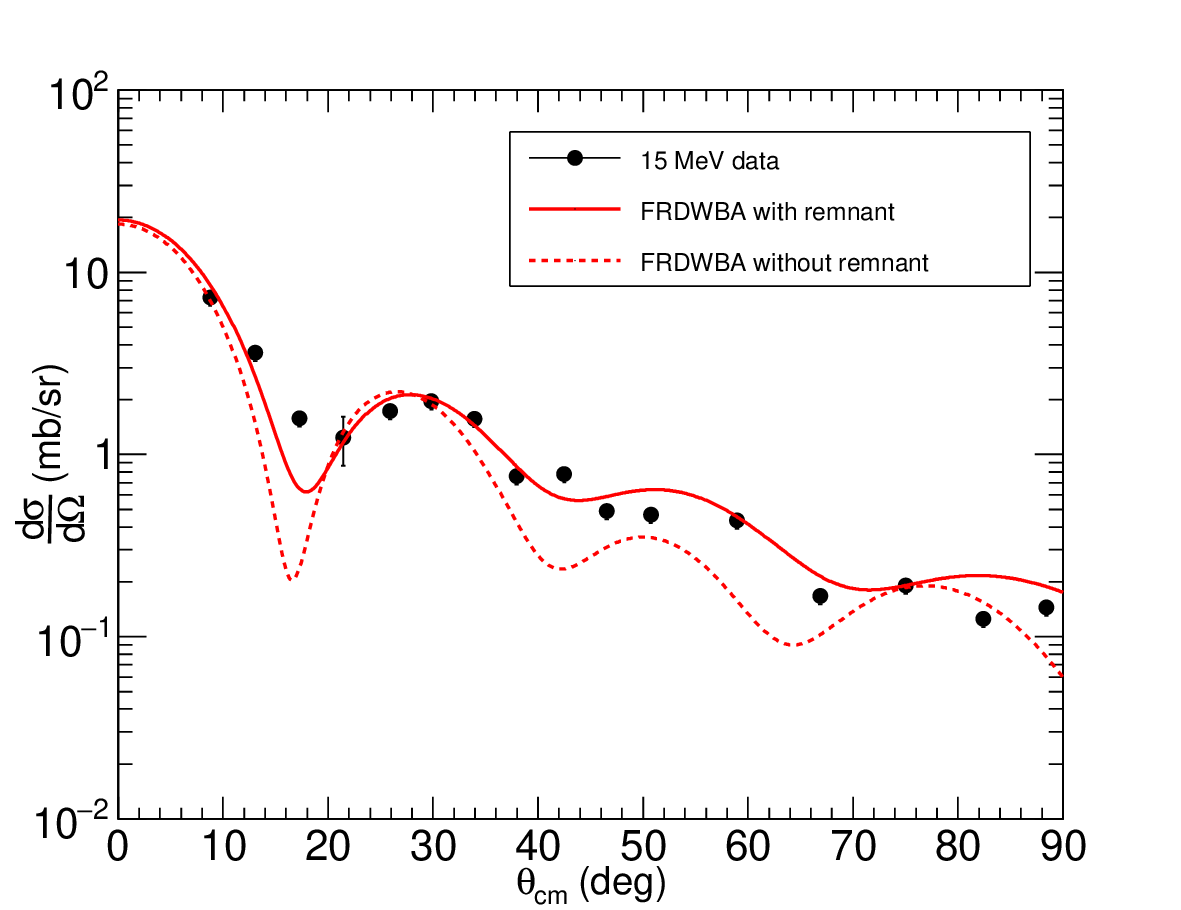}
    \caption{Angular distributions of the 8664 keV state from the $^{22}$Ne($^3$He,$d$)$^{23}$Na  reaction at 15 MeV~\cite{PO71}. The FRDWBA calculations are shown by the solid and dotted lines.}
    \label{fig : xs}
\end{figure}
 The real binding potentials for the ($d$+$p$) and $^{22}$Ne+$p$ systems were also included, with their depths adjusted to reproduce the effective proton separation energy. The potentials were of the standard Woods-Saxon shape. The potential parameters are listed in Table~\ref{tab:OMP}. The code FRESCO~\cite{TH88} was used to carry out the calculations. The resultant DWBA calculations along with the data are shown in Fig.~\ref{fig : xs}. The proton spectroscopic factors $S$, were extracted by normalizing the calculated DWBA calculations to the experimental data,
 \begin{align}
     \bigg(\frac{d\sigma}{d\Omega}\bigg)_\textrm{Exp}=S \bigg(\frac{d\sigma}{d\Omega}\bigg)_\textrm{DWBA}
 \end{align}
 
The spectroscopic factor for $^3$He in $(d+p)$ configuration is taken as 1.16~\cite{KI20}. The proton spectroscopic factors for the 8664 keV state from the present calculations are relatively higher than those obtained in Ref.~\cite{DU67,PO71} (Table~\ref{tab:ANC}). Note that in these previous works, the zero-range DWBA calculations used a normalizing factor of 4.42 to explain the experimental data~\cite{DU67,PO71}. 
Inclusion of the complex remnant term in the present FRDWBA calculations results in a better overall fit to the data, as also seen by Ref.~\cite{SA20}. The dotted lines in Fig.~\ref{fig : xs} represent the FRDWBA calculation sans the remnant term.
\begin{table*}[th!]
\caption{\label{tab:OMP}Potential parameters used in the present work. $V$ and $W$ are the real and imaginary depths in MeV, $r$ and $a$ are the radius and diffuseness in fm. $R_x = r_x A^{1/3}$ fm ($x=V, W, S, SO, C$).}
\begin{ruledtabular}
\begin{tabular}{c  c  c c c c   c  c  c  c  c  c  c  c  c}
Channel &$V$ &$r_{V}$ &$a_{V}$ &$W_{V}$ &$r_{W}$ &$a_{W}$ &$W_{S}$ &$r_{S}$ &$a_{S}$ &$V_{SO}$ &$r_{SO}$ &$a_{SO}$ &$r_C$  &Ref.\\
\hline 
$^{22}$Ne + $^3$He &177.0 &1.14   &0.72 &$13.0$ &$1.60$ &$0.77$    &$-$  &$-$   &$-$  &$8.0$    &$1.14$  &$0.72$ &1.40     &\cite{PO71}\\
$d$ + $^{23}$Na &105.0  &1.02   &0.86 &$-$ &$-$ &$-$    &$80.0$  &$1.42$   &$0.65$  &$6.0$  &$1.02$  &$0.86$ &1.30   &\cite{PO71}\\
$d$ + $^{22}$Ne   &88.0  &1.17   &0.73 &0.24 &1.33 &0.73   &$35.8$  &$1.33$    &$0.73$    &13.85  &1.07    &0.66  &1.33  &\cite{SA20}\\
$d$ + $p$ &\footnote{varied to match separation energy} &1.25 &0.65 &$-$ &$-$ &$-$  &$-$  &$-$   &$-$  &$6.2$  &$1.25$   &$0.65$ &1.30 &\cite{SA20}\\
$p$ + $^{22}$Ne   &\footnotemark[1]  &1.26   &0.60 &$-$ &$-$ &$-$  &$-$  &$-$   &$-$    &6.2  &1.26    &0.60  &1.33  &\cite{SA20}\\
\end{tabular}
\end{ruledtabular}
\end{table*}
\begin{table*}[th!]
\caption{\label{tab:ANC}Spectroscopic factors ($S$) and ANC ($C$) of $E_{x}=7080,7449,7890$ and 8664 keV state of $^{23}$Na from the present work.  }
\begin{ruledtabular}
\begin{tabular}{ c c c c c c c c } 
E$_x$  &{$J^\pi$}  &{${nl}_j$}  &{$S$} &{$S$}  &$b$ (fm$^{-1/2}$) &$C$ (fm$^{-1/2}$) &$C$ (fm$^{-1/2}$)  \\ 
(keV)  &  &  &(Present) &(Literature) &(Present) &(Present) &(Literature)  \\ 
\hline                    
8664 &${1/2}^{+}$ &$2s_{1/2}$  &$0.50 \pm 0.05$ &$0.32 \pm 0.05$~\cite{SA20} &249 &$179.5 \pm  19.7$ &$143.7 \pm 15.2$~\cite{SA20}\\
     &            &             & &$0.58 \pm 0.08$~\cite{TE93} & & &\\
       &   &   &   &$0.42 \pm 0.08$~\cite{FE18} & & & \\
       &   &   &   &$0.30$~\cite{GO82} & & & \\
       &   &   &   &$0.29$~\cite{HA01} & & & \\
       &   &   &   &$0.31$~\cite{DU67} & & & \\
       &   &   &   &$0.27$~\cite{PO71} & & & \\
\hline 
7080 &${1/2}^{-}$ &$2p_{1/2}$   & $0.08 \pm 0.01$ &0.3~\cite{DU67}    &8.95     &$2.53 \pm 0.16$          &                   \\
     &             &            &       &0.085~\cite{PO71}  &   &   &\\\\
7449 &${3/2}^{+}$ &$1d_{3/2}$   & $0.06 \pm 0.01$  &0.28~\cite{DU67}   &3.93     &$0.96 \pm 0.06$          &                   \\
 &  &   &   &0.14~\cite{PO71} & &   &\\\\
7890 &${3/2}^{+}$ &$1d_{3/2}$   & $0.05 \pm 0.01$  &0.15~\cite{DU67}  &3.75     &$0.84 \pm 0.08$          &                   \\
    &   &   &   &0.11~\cite{PO71} & &   &\\
\end{tabular}
\end{ruledtabular}
\end{table*}
\begin{table*}[htb!]
  \centering
  \caption{Background pole parameters obtained from $R-$matrix fits.}
  \label{tab:poles}
  \begin{tabularx}{\textwidth}{c c c X X X X X X X}
    \hline
    \hline
    \multirow{2}{*}{$J^\pi$} & \multirow{2}{*}{E$_x$ (MeV)} & \multirow{2}{*}{$\Gamma_p$ (MeV)} & & &\multicolumn{4}{c}{$\Gamma_\gamma[E_1]$ (eV)} &  \\
    \cline{4-10}
    & &  & $R\rightarrow \textrm{g.s}$ & $R\rightarrow 0.44$ & $R\rightarrow 2.39$ & $R\rightarrow 2.98$ & $R\rightarrow 6.30$ &$R\rightarrow 6.91$ &$R\rightarrow 8.66$ \\
    \hline
    $1/2^{-}$ & 15 & 5.0 & 822.91 & &$3.17\times10^3$ & 697.67  &$1.58\times10^3$  &  & 139.528 \\
    $3/2^{-}$ & 15 & 5.0 & 20.92 & 974.42 &  &  & &  &  \\
    \hline
    \hline
  \end{tabularx}
\end{table*}

As discussed in Ref.~\cite{SA20}, the spectroscopic factors are dependent on the choice of potential parameters and are sensitive to the geometric parameters of the bound state potentials. 
Thus, instead of spectroscopic factor, the ANC is a more appropriate quantity. The ANC method is free from the geometrical parameters of the binding potentials and relies primarily on the peripheral nature of the reaction. The square of the ANC ($C^2$) of a particular state is related to the spectroscopic factor ($S$) via the single particle ANC ($b$) as
\begin{align}
    C^2=Sb^2 \label{eq : ANC}
\end{align}
The single particle ANC ($b$) is the normalization of the bound state wave function of the composite nucleus $^{23}$Na at large radii with respect to the Whittaker function.
\begin{figure}[h!]
    \centering 
    \includegraphics[width =0.5\textwidth]{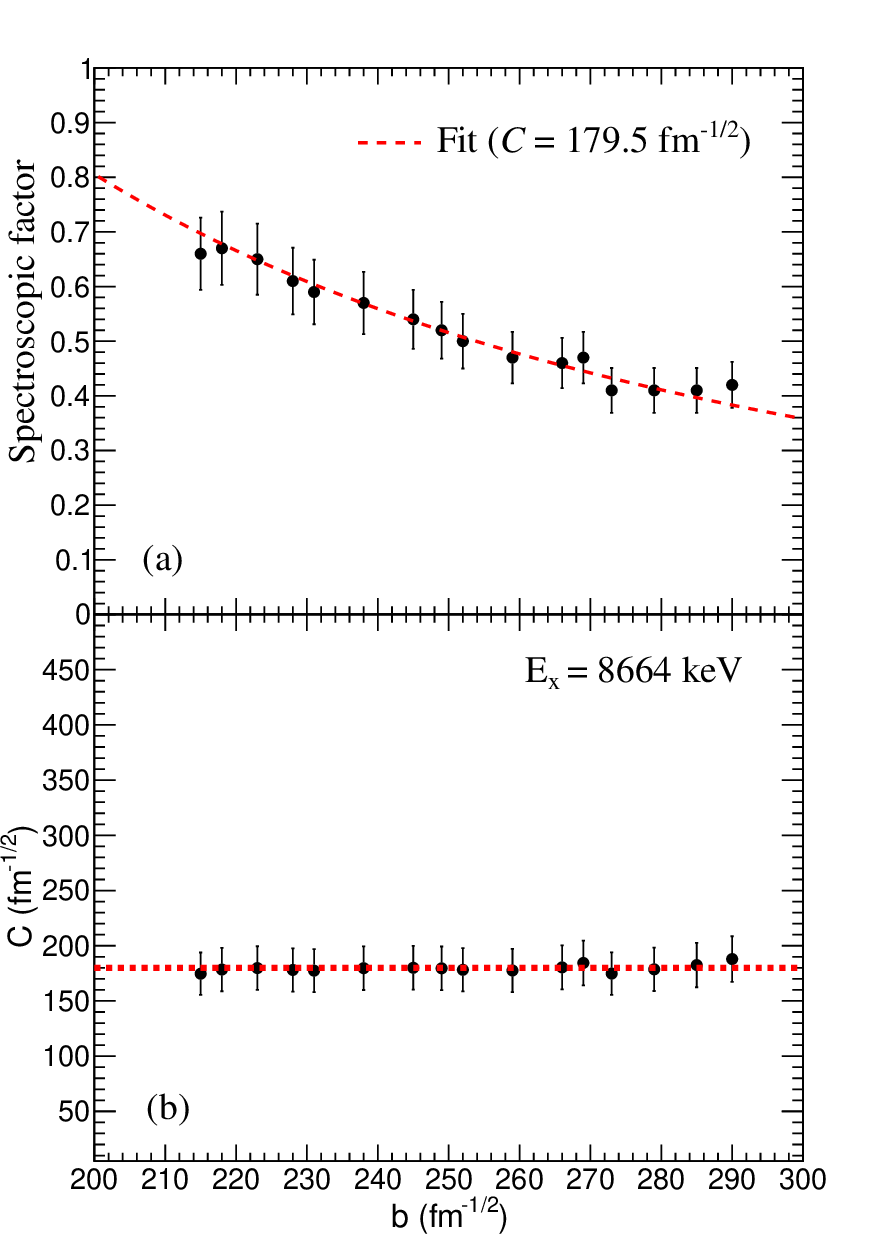}
    \caption{(a) Variation of spectroscopic factor ($S$) with single particle ANC ($b$) for the 8664 keV state. (b) Variation of ANC ($C$) with single particle ANC ($b$) for the 8664 keV state. }
    \label{fig : S vs b}
\end{figure}
\begin{figure}[h!]
    \centering 
    \includegraphics[width =0.5\textwidth]{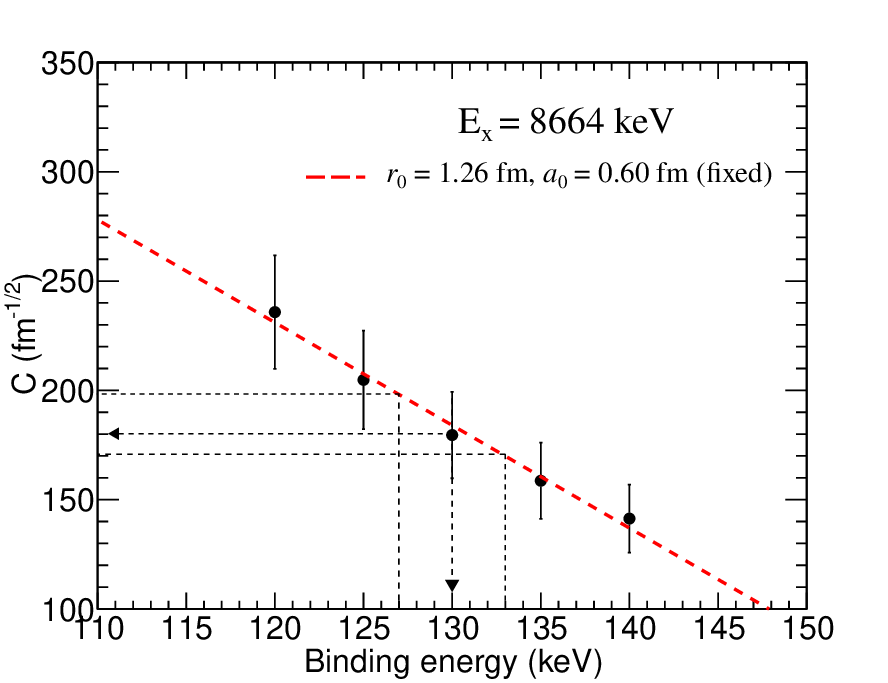}
    \caption{Variation of ANC ($C$) with binding energy for the 8664 keV state. }
    \label{fig : C vs BE}
\end{figure}

To test the peripheral condition, the variation of the spectroscopic factor ($S$) against the single particle ANC ($b$) was studied for the 15 MeV data as shown in Fig.~\ref{fig : S vs b}~(a). The single particle ANC ($b$) is varied by changing the geometrical parameters of the $^{22}$Ne + $p$ binding potential in small steps.
According to Eq.~\ref{eq : ANC}, the variation of $S$ should be proportional to the inverse
square of $b$. From Fig.~\ref{fig : S vs b}(a), the $S$ obtained from the 15 MeV data follows the inverse square relation. Hence, the ANC extracted from the 15 MeV data is peripheral and this ANC is considered for all further calculations.
In Fig.~\ref{fig : S vs b} (b), the extracted ANC is plotted as a function of $b$. The mean ANC obtained is 179.5~fm$^{-1/2}$ and is shown with the dotted line. In Fig.~\ref{fig : C vs BE}, the 
dependence of the ANC as a function of binding energy for the 8664 keV state is shown. 
The binding energy for the 8664 keV is $130 \pm 3$ keV~\cite{SA20}. It is varied by keeping the geometry parameters of the bound state potential fixed ($r_0=1.26$~fm, $a_0=0.60$~fm) corresponding to the mean ANC value. The plots show that the ANC value of the 8664 keV subthreshold state decreases with increasing binding energy. The ANC value for the 8664 keV state from the present work is $\sim 25\%$ higher compared to that in Ref.~\cite{SA20}.

The uncertainty of the mean value of ANC has contribution from two sources. The first one arises due to the propagation of the error of spectroscopic factors ($S$)
through the relation given in Eq.~\ref{eq : ANC}. The errors in the values of $S$ include the uncertainty of experimental angular distribution data. Secondly, the contribution of the uncertainty in binding energy is added in quadrature to obtain the total uncertainty in the mean ANC value. The dotted lines in Fig.~\ref{fig : C vs BE} shows the variation in ANC due to the $\pm 3$ keV uncertainty in binding energy of the 8664 keV state.

Similar analyses were carried out to obtain the ANC values for the 7080, 7449 and 7890 keV states using the 15 MeV data which were not considered in the earlier work by Santra~\textit{et~al.}~\cite{SA20}. The extracted spectroscopic factors and ANCs from the present work along with the values available in the literature are listed in Table~\ref{tab:ANC}. 
\begin{figure}[h!]
    \centering 
    \includegraphics[width =0.5\textwidth]{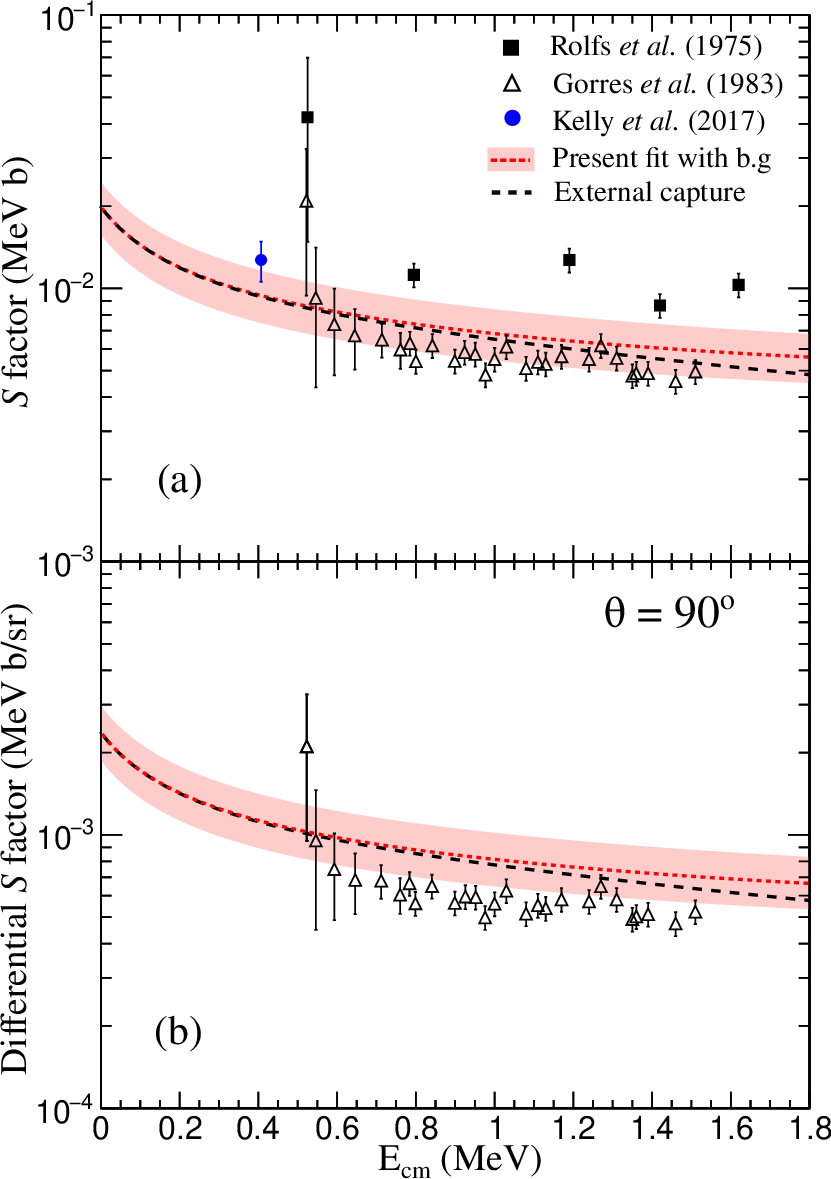}
    \caption{(a) Astrophysical $S-$factor and (b) differential $S-$factor at $\theta=90^\circ$, obtained from the $R-$matrix calculations for the DC $\rightarrow$ 8664 keV transition.  The red dotted line corresponds to the $S-$factor using the mean ANC of $179.5$~fm$^{-1/2}$ for the 8664 keV state. The bands correspond to error in $S-$factor due to uncertainty in the ANC value.}
    \label{fig : S-factor 8664}
\end{figure}

\subsection{\textit{R}-matrix calculations}
In the present work, a phenomenological $R-$matrix analysis was performed using the code \texttt{AZURE2}~\cite{AZ10}. The basic $R$-matrix theory used in the code \texttt{AZURE2} is described in Ref. \cite{AZ10, Lane}. In the $R$-matrix calculations, the channel radius ($r_c$) divides the radial space into internal and external parts.
For the present calculations, $r_c = 5.3$~fm, is obtained for the $^{22}$Ne + $p$ system by $\chi^2$ minimization procedure employing a grid search technique, keeping the ANCs fixed. The search was performed on the total non-resonant $S-$factor data to choose the $r_c$.
The $R$-matrix fitting has been performed on the available capture data of individual states and total non-resonance capture data simultaneously. The ANCs for the ground and the first five bound excited states were taken from Santra et al.~\cite{SA20}. 
For the three bound excited states at 7080, 7449, 7890 keV and the 8664 keV subthreshold state, the ANC values are taken from this work.
Two background poles with spin parity 1/2$^-$ and 3/2$^-$ are included, and only the $E1$ decay was considered to simulate the internal capture part of present $R$-matrix calculations. The excitation energies of the poles are chosen at 15 MeV and proton partial widths are fixed at 5 MeV, calculated from Wigner limit approximations. However, the gamma decay partial widths of the background poles are left as free parameters, with the initial values taken from the Weisskopf limit for the corresponding gamma transitions. The fitted background pole parameters are shown in Table \ref{tab:poles}.  

The present $R$-matrix modelling have two parts, at first the DC$\rightarrow$8664 keV  calculations are carried out, the resultant $S-$factor is compared with the existing data. The next part consists of the calculations for the DC$\rightarrow$GS transition and the total direct capture contribution. 
In order to see the effect of the enhanced ANC of the 8664 keV state, first the $R-$matrix calculations for DC$\rightarrow$8664 keV transition is carried out. The resultant $S-$factor and the differential $S-$factor at $\theta =90^\circ$ obtained using the mean peripheral ANC of $179.5$~fm$^{-1/2}$ are shown with the red dotted lines in Fig.~\ref{fig : S-factor 8664}(a) and (b), respectively. The band corresponds to the error in $S-$factor due to the $\sim 11\%$ uncertainty in the ANC value. The $S-$factor obtained by including the peripheral ANC from this work is able to explain the DC$\rightarrow$8664 keV capture data by Gorres~\textit{et~al.}~\cite{GO82} (open black triangles) within the uncertainty band.
 However, the data points by Rolfs~\textit{et~al.}~\cite{RO75} (filled black squares) and Kelly~\textit{et~al.}~\cite{KE17} (filled blue circle) lie above the maximum limit of the $S-$factor band.

\begin{figure}[h!]
    \centering 
    \includegraphics[width =0.5\textwidth]{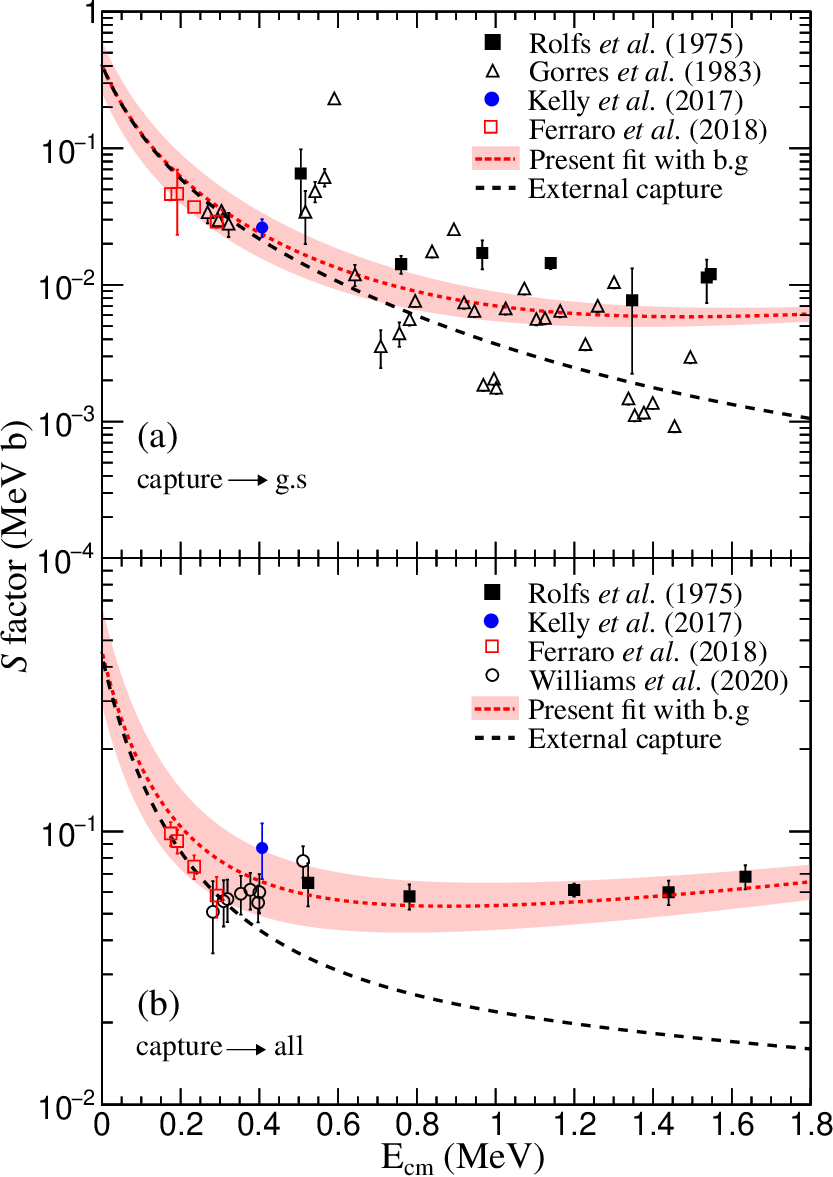}
    \caption{Astrophysical $S-$factor of the non-resonant capture in $^{22}$Ne($p,\gamma$)$^{23}$Na from previous direct measurements~\cite{RO75,GO82,KE17,FE18,WI20} and present $R-$matrix calculations. (a) Capture to the ground state in $^{23}$Na, (b) total $S-$factor. (see text for details)}
    \label{fig : S-factor}
\end{figure}

The 8664 keV state has a lifetime of $0.14 \pm 0.03$ fs and it decays to the ground state with a branching of $(84 \pm 3)\%$ ($\Gamma_\gamma = 4.7 \pm 1$ eV)~\cite{GO82}. The DC$\rightarrow$GS transition is influenced by the high energy tail of this $s-$wave subthreshold resonance ($E_p = - 130$ keV) as evident from the rise in the low energy $S-$factor data of Gorres~\textit{et~al.}~\cite{GO82} and Ferraro~\textit{et~al.}~\cite{FE18} (red squares) in Fig.~\ref{fig : S-factor}(a). The corresponding $S-$factor for the DC$\rightarrow$GS  
transition from the $R-$matrix calculations are shown with the red dotted lines in Fig.~\ref{fig : S-factor}(a). The calculations are in very good agreement with the low energy data (E$_\textrm{cm} < 400$ keV) of Gorres~\textit{et~al.}~\cite{GO82}. The data of Ferraro~\textit{et~al.}~\cite{FE18} and Kelly~\textit{et~al.}~\cite{KE17} lie inside the uncertainty band. At higher energies, the calculated $S-$factor passes through the data points of Gorres~\textit{et~al.}~\cite{GO82} but underestimates the data of Rolfs~\textit{et~al.}~\cite{RO75}.
The rising nature of the $S-$factor at low energies is very well reproduced by the calculations.

The total $S-$factor for the non-resonant capture in $^{22}$Ne($p$,$\gamma$)$^{23}$Na reaction is obtained by adding the $S$-factors of all the individual transitions to the ground and the bound excited states is shown with the red dotted line in Fig.~\ref{fig : S-factor}(b). The calculations were carried out by including the ANCs for the three bound excited states 7080, 7449 and 7890 keV and the 8664 keV subthreshold state obtained from this work (Table~\ref{tab:ANC}).
The ANCs for the ground and the first five bound excited states were taken from Ref.~\cite{SA20}.  The bands correspond to error in $S-$factor due to uncertainty in the ANC values and uncertainty in the decay width of the 8664 keV state.
The total $S-$factor from the calculations is in excellent agreement with the recent direct measurement data of Williams~\textit{et~al.}~\cite{WI20} as well as with the lower energy data of Ferraro~\textit{et~al.}~\cite{FE18}.

\subsection{Shell model calculations and partial widths for the 8945 keV resonance}
The extraction of spectroscopic factors for the 8945 keV doublet by Hale \textit{et al}.~\cite{HA01} and Santra \textit{et al}.~\cite{SA20} is not completely reliable due to the scarce angular distribution data. In this work, the proton spectroscopic factor for the 7/2$^{-}_{2}$ state of $^{23}$Na at 8945 keV has been calculated using the NUSHELLX code~\cite{nushellx}. Large basis shell model (LBSM) calculations were performed. The positive parity states were easily be reproduced using the \textit{sd} model space. But, for the negative parity state, the upper \textit{pf} shell were taken with the \textit{sd} shell. Thus, the \textit{sdpf} model space was used to get  the negative parity states. The full model space calculation is constrained due to the present computational capacity. Hence, a suitable truncation scheme was adopted. 
Subshell restrictions were chosen with zero occupancy in the $1f_{5/2}$ and $2p_{1/2}$ subshells. The \textit {sdpfmu} interaction~\cite{sdpfmu} with the mentioned truncation scheme reproduces the experimentally observed 7/2$^{-}_{2}$ state of $^{23}$Na at 9173 keV. The calculated energy level is 228 keV above the experimentally adopted energy level. Similarly, the 3/2$^{+}_{8}$ doublet state has been reproduced theoretically at 9023 keV energy. 

The single proton spectroscopic factor is calculated for the astrophysically important 7/2$^{-}_{2}$ and 3/2$^{+}_{8}$ doublet states of $^{23}$Na at 8945 and 8944 keV. The proton spectroscopic factor for the 3/2$^{+}_{8}$ state from the shell model calculations is $1\times 10^{-4}$ which is consistent with the order of the experimental values. For the 7/2$^{-}_{2}$ state, the value obtained from the calculation is 0.0104 which is substantially higher compared to the earlier studies~\cite{HA01,SA20}.
To validate the theoretical calculations, the spectroscopic factors for the low-lying states of $^{23}$Na were also calculated and compared with the corresponding experimental values as shown in Table~\ref{tab:SM-sf}.
The theoretical calculations are in good agreement with the experimentally determined values. This consistency provides strong confidence to our theoretically obtained spectroscopic factor for the 7/2$^{-}_{2}$ state.
\begin{table}[h!]
\centering
\caption{\label{tab:SM-sf}Comparison of spectroscopic factors ($S$) for the low-lying bound states and the 8945 keV doublet states of $^{23}$Na from the present shell model calculations ($S^\textrm{SM}$) and literature.}
\smallskip
\begin{tabular}{c c c c c}
\hline \hline 
E$_{x}$ &$J^{\pi}$ &${nl}_{j}$ &$S^\textrm{SM}$ &$S$\\
(keV) &          &       & (This work) & (Literature) \\
\hline
g.s &3/2$^{+}$ &1$d_{3/2}$ &0.055 &0.08~\cite{PO71}\\
    &          &           &      &$0.082 \pm 0.012$~\cite{SA20}\\\\  
440 & 5/2$^{+}$&1$d_{5/2}$&0.41&0.35~\cite{PO71}\\
    &          &          &    &$0.38 \pm 0.08$~\cite{SA20}\\\\
2392 &1/2$^{+}$&2$s_{1/2}$&0.20&0.25~\cite{PO71}\\
    &           &           &   &$0.26 \pm 0.05$~\cite{SA20}\\\\
2982 &3/2$^{+}$&1$d_{3/2}$&0.23&0.32~\cite{PO71}\\
    &           &       &       &$0.35 \pm 0.04$~\cite{SA20}\\\\
6308 &1/2$^{+}$&2$s_{1/2}$&0.10&0.13~\cite{PO71}\\
    &           &       &       &$0.14 \pm 0.02$~\cite{SA20}\\\\
8945 &3/2$^{+}$&1$d_{3/2}$&1$\times 10^{-4}$& $(5.54 \pm 1.41)\times 10^{-4}$~\cite{SA20}\\
&&&&$8.32\times 10^{-4}$~\cite{HA01}\\\\
8944 &7/2$^{-}$&1$f_{7/2}$&0.0104&$(3.94\pm 0.9)\times 10^{-4}$~\cite{SA20}\\
&&&&$\leq$1.08$\times 10^{-3}$~\cite{HA01}\\
\hline
\end{tabular}
\end{table}
\begin{table}
\centering
\caption{Resonance strengths of E$_x=8945$ keV doublets (3/2$^+$, 7/2$^-$) from the Wigner limit and present shell model calculations, compared with previous experimental measurements.}
\label{tab:resonance strength}
\begin{tabular}{cccccccccccc}     \hline \hline
E$_x$ &E$_\textrm{cm}$    &J$^\pi$ &$\omega\gamma^\textrm{UL}$ &$\omega\gamma^\textrm{SM}$ &$\omega\gamma^\textrm{Exp.}$& \\ 
(keV) &(keV)  &  &(eV)&(eV)  &(eV) \\ \hline
8945 &151  &3/2$^+$&1.17$\times$10$^{-7}$&3.6$\times$10$^{-8}$ & $(1.9\pm0.1)\times$10$^{-7}$\cite{WI20} \\
& & & & &$(1.48 \pm 0.1)\times$10$^{-7}$\cite{DE16}& \\
& & & & &$(1.8 \pm 0.2)\times$10$^{-7}$\cite{CA18}& \\
& & & & &$(2.2\pm 0.2)\times$10$^{-7}$\cite{FE18}& \\
& & & & &2.03(40)$\times$10$^{-7}$\cite{KE17}& \\
& & & & &$(2.0\pm 0.5)\times$10$^{-7}$\cite{SA20}& \\
& & & & &1.7$^{+0.50}_{-0.40}\times$10$^{-7}$\cite{LE20}& \\
& & & & & & \\ 
8944 &150 &7/2$^-$&1.89$\times$10$^{-10}$& 9.97$\times$10$^{-8}$&$\leq$9.2$\times$10$^{-9}$\cite{HA01} \\
&&&&&$\leq$9.7$\times$10$^{-8}$\cite{KE17} \\
&&&&&$(3.93\pm 0.9)\times$10$^{-9}$\cite{SA20} \\ \hline
\end{tabular}
\end{table}

The proton partial widths ($\Gamma_p$) for the 8945 keV doublet is obtained using the relation
\begin{align}
    \Gamma_p =S\Gamma_{sp}
\end{align} 
where $\Gamma_{sp}$ is the single particle width of a resonance for pure single particle configuration calculated from the code DWUCK4~\cite{Kunz} and $S$ is the spectroscopic factor from shell model calculations. For comparison, the proton partial widths using the Wigner limit is also calculated. The values of $\Gamma_{sp}$ for the $3/2^{+}$ and $7/2^{-}$ states are $1.8 \times 10^{-4}$ eV and $2.49\times10^{-6}$ eV respectively. 
The proton partial width ($\Gamma_p$) can be expressed as the product of an energy-dependent penetration factor, $P_l(E)$ and an energy independent reduced width, $\gamma_p^2$, as
\begin{align}
    \Gamma_p = 2P_l(E)\gamma_p^2
\end{align} 
where,
\begin{align}
    \gamma_p^2=\frac{\hbar^2}{\mu a^2}\theta_p^2
\end{align}
The constant $\hbar^2/(\mu a^2)$ is the Wigner limit, where $a$ is the channel radius, $\mu$ is the reduced mass and $\theta_p^2$ is proportional to the proton spectroscopic factor. 
The partial gamma decay widths ($\Gamma_\gamma$) for $M1$ and $E1$ transitions obtained from the Weisskopf estimates~\cite{IL07} are 0.37~eV and 133.62~eV, respectively. The $7/2^{-}$ state of the 8945 keV doublet undergoes decay to ${9/2}^{-}$ and ${9/2}^{+}$ states emitting $\gamma$-rays with energies 2592 keV($M1$) and 6240 keV($E1$), respectively~\cite{JE13}.

\section{Thermonuclear reaction rate of $^{22}$N\lowercase{e}($p,\gamma$)$^{23}$N\lowercase{a}}
The thermonuclear reaction rate of $^{22}$Ne($p$,$\gamma$)$^{23}$Na reaction is governed by various low energy narrow resonances and the non-resonant component. As there are no interfering resonances to consider for the $^{22}$Ne($p$,$\gamma$)$^{23}$Na reaction, the total resonant rate $ N_A<\sigma v>_R$ is given by the sum of all narrow resonances~\cite{DE16},
\begin{align}
    N_A<\sigma v>_R=\frac{1.5399 \times 10^5}{(\mu T_9)^{3/2}}\sum_{i} (\omega \gamma)_i e^{\frac{- 11.605\textrm{E}_{\textrm{cm},i}}{T_9}}
\end{align}
where $T_9$ is the temperature in GK, $\mu$ is the reduced mass in amu, $(\omega \gamma)_i$ the strength of resonance $i$ in eV, and $\textrm{E}_{\textrm{cm},i}$ is the center-of-mass energy of resonance $i$ in MeV. 

The resonance strengths for the 8945 keV doublet is derived using the relation,
\begin{align}
    \omega \gamma = \frac{2J+1}{(2j_1+1)(2j_2+1)}(1+\delta_{12})\frac{\Gamma_p\Gamma_\gamma}{\Gamma}
\end{align}
where $j_1$ and $j_2$ are the spins of the interacting particles i.e., $^{22}$Ne and $p$, and $J$ is the spin of the excited state populated in the compound nucleus i.e., $^{23}$Na. $\Gamma$ is the total width i.e., $\Gamma_p + \Gamma_\gamma$. 
The values of the resonance strengths for the 3/2$^+$ and $7/2^{-}$ state obtained from the present shell model calculations ($\omega \gamma^\textrm{SM}$) and Wigner limit ($\omega \gamma^\textrm{UL}$) calculations are compared with the corresponding experimental values ($\omega \gamma^\textrm{Exp.}$) in Table~\ref{tab:resonance strength}. Unlike the $7/2^{-}$ state, the strength of the $3/2^{+}$ state has been very well constrained by various measurements (Table~\ref{tab:resonance strength}). In this work, we use the strength value adopted by Williams et al.~\cite{WI20} for the $3/2^{+}$ state and the strength of the $7/2^{-}$ state is determined from the present shell model calculations. Recently, a new experimental study of the $^{23}$Na+$p$ inelastic-scattering reaction at the Q3D magnetic spectrometer at Munich has ruled out the earlier reported resonances at E$_x=8862$, $8894$ and $9000$ keV~\cite{AD23,CA23}. Hence, these resonances are omitted from the present work. For resonances at E$_\textrm{cm}=35$, 178, 417, 458, 610, 632 and 1222 keV, the strength values are also taken from Ref.~\cite{WI20}. The strengths of resonances located between 632 and 1222 keV, and beyond 1222 keV, are adopted from Ref.~\cite{SA13}. The strength values were further divided by the corresponding electron screening enhancement factor, taken from Ref.~\cite{WI20}.
\begin{figure}[h!]
    \centering 
    \includegraphics[width =0.5\textwidth]{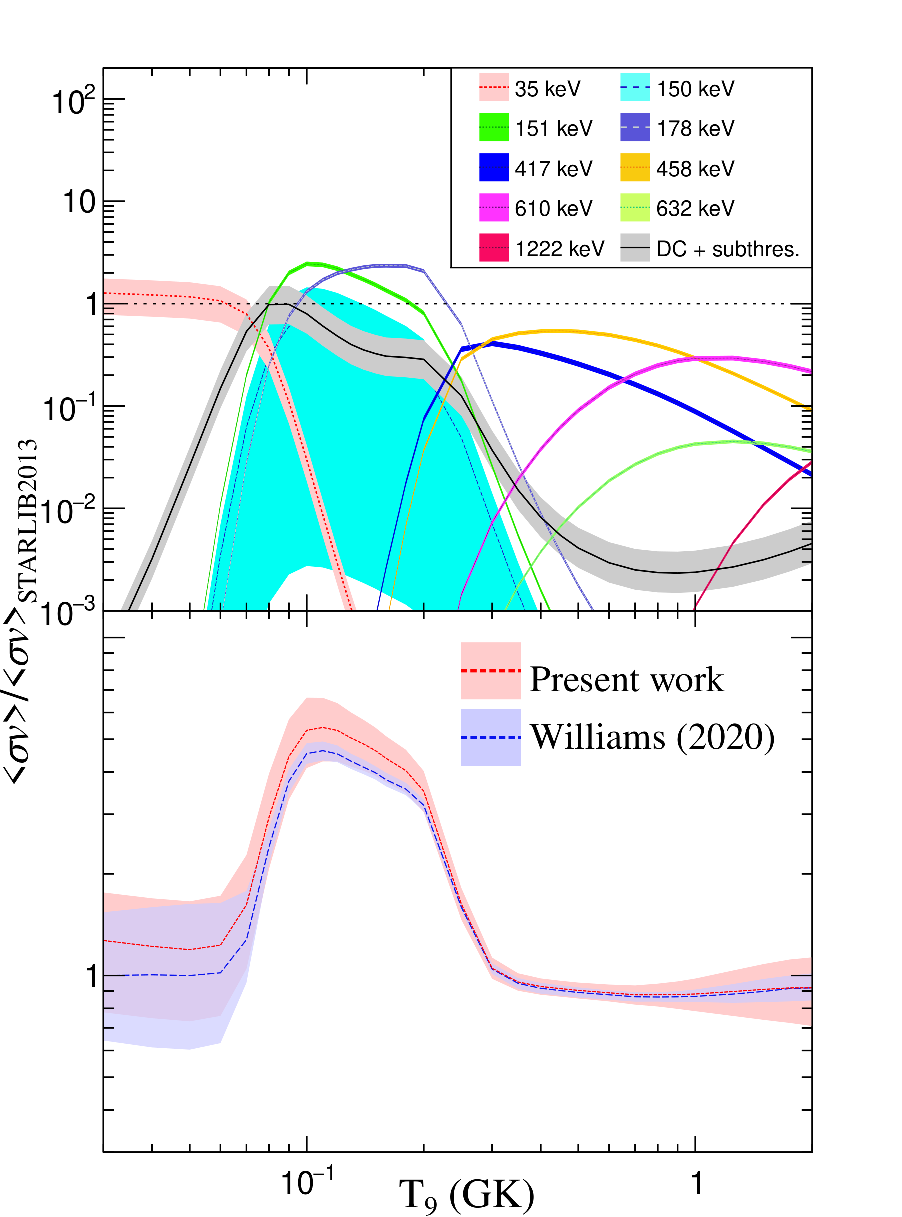}
    \caption{Reaction rate for the $^{22}$Ne($p,\gamma$)$^{23}$Na reaction as a function of temperature in GK. (a) Comparison of the individual resonant and direct capture contributions relative to the \texttt{STARLIB2013} rate. (b) The total rate from the present work (red) and from previous measurement (blue) by Williams et al.~\cite{WI20} normalized to the \texttt{STARLIB2013} rate. The bands correspond to the uncertainties in the rates and the dashed lines represent the mean rates. }
    \label{fig : rate}
\end{figure}

The contributions of various individual resonances and the non-resonant component (DC+subthres.) normalized to the median \texttt{STARLIB-2013} rates~\cite{SA13} are shown in the top panel of Fig.~\ref{fig : rate}. At very low temperatures, $T_9 \leq 0.05$, the reaction rate is dominated by the 35 keV resonance. The temperature range $T_9 = 0.08 - 0.1$ is significant for the process of hot bottom burning (HBB) in asymptotic giant branch (AGB) stars~\cite{FE18}. In the previous studies by Ferraro et al.~\cite{FE18} and Santra et al.~\cite{SA20}, the 68 and 100 keV resonance had large contributions at these temperatures. However, in this work, these resonances are removed, and the rate is affected by the doublets at 150 keV, the resonance at 178 keV and the non-resonant capture component.
The non-resonant component is obtained from the present $R$-matrix calculations of the $S$-factor due to the DC$\rightarrow$8664 keV transition. The non-resonant rate from the present work is $\sim 30\%$ higher at $T_9 = 0.05–0.1$ compared to the earlier rates of Depalo et al.~\cite{DE16} and Kelly et al.~\cite{KE17}, and slightly higher relative to the rates of Ferraro et al.~\cite{FE18} and Santra et al.~\cite{SA20}.  

The bands in Fig.~\ref{fig : rate} correspond to the uncertainties in the rates, and the lines represent the mean rates. The uncertainties in the individual resonant rates are primarily due to the uncertainties in the resonance strength values, whereas for the non-resonant component, the major contribution is the uncertainty in the $S$-factor. The uncertainty in the $7/2^{-}$ doublet at 150 keV is shown with the cyan band. The upper and lower limits of the rate for the $7/2^{-}$ state are due to the strength values from the present shell model calculations and Wigner limit, respectively (Table~\ref{tab:resonance strength}). The mean value (dotted lines in blue) is obtained by taking the average of the upper and lower limits.

The total reaction rate is obtained by adding the resonant and non-resonant capture contributions. In the bottom panel of Fig.~\ref{fig : rate}, the normalized total rate from the present work (red) is compared to the rate by Williams et al.~\cite{WI20} (blue). The rate from Williams et al.~\cite{WI20} lies within the uncertainty limits of the rate from the present work.
At low temperatures $T_9 \leq 0.05$, and at slightly higher temperatures, $T_9 =0.08-0.1$, relevant for the AGB stars, our mean rate (red dashed lines) is $\sim 15\%$ higher than the mean rate of Williams et al.~\cite{WI20} (blue dashed lines). For high temperatures $T_9 = 0.2-0.25$, responsible for classical novae nucleosynthesis, our mean rate coincides with the mean rate of Williams et al.~\cite{WI20}. 

\section{Conclusion}
The $^{22}$Ne($p$,$\gamma$)$^{23}$Na reaction is reanalysed by extracting the ANC of the 8664 keV subthreshold state from finite-range DWBA analysis of the existing transfer reaction data of $^{22}$Ne($^3$He,$d$)$^{23}$Na at 12 and 15 MeV~\cite{DU67,PO71}. The contributions of the previously neglected bound excited states at 7080, 7449 and 7890 keV are also included in the present work. ANC value of $217\pm 38.6$ fm$^{-1/2}$ for the 8664 keV state, obtained from the 12 MeV data, satisfied the necessary peripherality checks and is further utilized to carry out $R$-matrix calculations. 
The astrophysical $S$-factor for the DC$\rightarrow$8664 keV using the enhanced ANC value explains the existing data of Rolfs et al.~\cite{RO75} and Kelly et al.~\cite{KE17} but overestimates the data of Gorres et al.~\cite{GO82}. The observed rise in the $S$-factor of the capture to ground state at low energies is reproduced nicely without requiring any background poles fitting.  The total non-resonant $S$-factor from the present work is in good agreement with the measurements by Ferraro et al.~\cite{FE18} and Williams et al.~\cite{WI20}.

The proton partial widths for the 8945 keV doublets ($3/2^{+}$ and $7/2^{-}$) are deduced from shell model calculations with the code NUSHELLX and are compared with the widths from Wigner limit. The resonance strength of the $3/2^{+}$ state is very well constrained from several experimental measurements, the value adopted by Williams et al.~\cite{WI20} is used in the present work. For the poorly studied $7/2^{-}$ state, the strength yielded from shell model and Wigner limit calculations have been used. 

The thermonuclear reaction rate evaluated in this work omits the resonances at E$_x = 8862$, 8894 and 9000 keV. The total reaction rate normalized to the \texttt{STARLIB-2013} rate~\cite{SA13} is compared to the rate of Williams et al.~\cite{WI20}. At the temperature range of interest for HBB processes ($T_9=0.08-1$), the mean rate from present work is $\sim 15\%$ higher than that by Williams et al.~\cite{WI20} and for higher temperatures relevant for classical novae nucleosynthesis ($T_9=0.2-0.25$), the present mean rate coincides with William's rate~\cite{WI20}.\\\\

\end{document}